\documentclass{appolb}
\usepackage{epsfig}
\begin{document}
\title{Multi-neutron transfer coupling in sub-barrier $^{32}$S + $^{90,96}$Zr 
fusion reactions}
\author{A. Richard$^{1}$,
C. Beck$^{1}$,
H.Q. Zhang$^{2}$,
C.J. Lin$^{2}$,
F. Yang$^{2}$,
H.M. Jia$^{2}$, 
X.X. Xu$^{2}$,
Z.D. Wu$^{2}$,
F. Jia$^{2}$,
S.T. Zhang$^{2}$,
Z.H. Liu$^{2}$
}
\address{ 
$^1$ Institut Pluridisciplinaire Hubert Curien, 
        UMR 7178, CNRS-IN2P3 and Universit\'{e} de Strasbourg, 
        B.P. 28, F-67037 Strasbourg Cedex 2, France \\
$^2$ China Institute of Atomic Energy (CIAE), 102413 Beijing, China
}
\maketitle

\begin{abstract}

\abstract{The role of neutron transfers is investigated in
the fusion process below the Coulomb barrier by analyzing 
$^{32}$S+$^{90}$Zr and $^{32}$S+$^{96}$Zr as benchmark 
reactions. A full coupled-channel calculation of the fusion excitation
functions has been performed for both systems by using multi-neutron 
transfer coupling for the more neutron-rich reaction. The enhancement 
of fusion cross sections for $^{32}$S+$^{96}$Zr is well reproduced at 
sub-barrier energies by NTFus code calculations including the coupling 
of the neutron-transfer channels following the Zagrebaev semiclassical 
model. We found similar effects for $^{40}$Ca+$^{90}$Zr and 
$^{40}$Ca+$^{96}$Zr fusion excitation functions.} 

\end{abstract}

%
%
\section{Introduction}
\label{intro}

Heavy-ion fusion reactions with colliding neutron-rich nuclei at 
bombarding energies at the vicinity and below the Coulomb barrier have 
been widely studied 
\cite{Dasgupta,Pengo83,Stelson88,Rowley92,Timmers98,Zagrebaev03,Stefanini06,Stefanini07,Yang08,Kalkal2010}.
The specific role of multi-step neutron-transfers in sub-barrier fusion 
enhancement still needs to be investigated in detail both experimentally 
\cite{Pengo83,Timmers98,Yang08} and theoretically \cite{Rowley92,Zagrebaev03}. 
In a complete description of the fusion dynamics the transfer channels in 
standard coupled-channel (CC) calculations 
\cite{Dasgupta,Rowley92,Zagrebaev03,Kalkal2010,CCFULL} have to be taken into 
account accurately. It is known that neutron transfers may induce a neck 
region of nuclear matter in-between the interacting nuclei favoring the 
fusion process to occur.

In low-energy fusion reactions, the very simple one-dimensional 
barrier-penetration model (1D-BPM) is based upon a real potential barrier
resulting from the attractive nuclear and repulsive Coulomb interactions.
For light- and medium-mass nuclei, one only assumes that the di-nuclear system
fuses as soon as it has reached the region inside the barrier i.e. within the 
potential pocket. If the system can evolve with a 
bombarding energy high enough to pass through the barrier and to reach this 
pocket with a reasonable amount of energy, the fusion process will occur 
after a complete amalgation of the colliding nuclei forming the compound 
nucleus. On the other hand, for sub-barrier 
energies the di-nuclear system has not enough energy to pass through 
the barrier.  In this case, neutron pick-up processes can occur 
when the nuclei are close enough to interact each other significantly 
\cite{Stelson88,Rowley92}, if the Q-values of neutron transfers 
are positive. 

It was shown that sequential neutron transfers can lead to the broad 
distributions characteristic of many experimental fusion cross sections. 
Finite Q-value effects can lead to neutron flow and a build up of a neck 
between the target and projectile \cite{Rowley92}. The situation
of this neck formation of neutron matter between the two colliding nuclei 
could be considered as a ``doorway state" to fusion. In a basic view, this 
intermediate state induced a barrier lowering. As a consequence, it will favor 
the fusion process at sub-barrier energies and enhance significantly the fusion 
cross sections. Experimental results have already shown such enhancement of 
the sub-barrier fusion cross sections due to the neutron-transfer channels 
with positive Q-values \cite{Pengo83,Timmers98}.

\section{Experimental results}
\label{sec:1}

\begin{figure}
\resizebox{0.95\columnwidth}{!}{%
\includegraphics{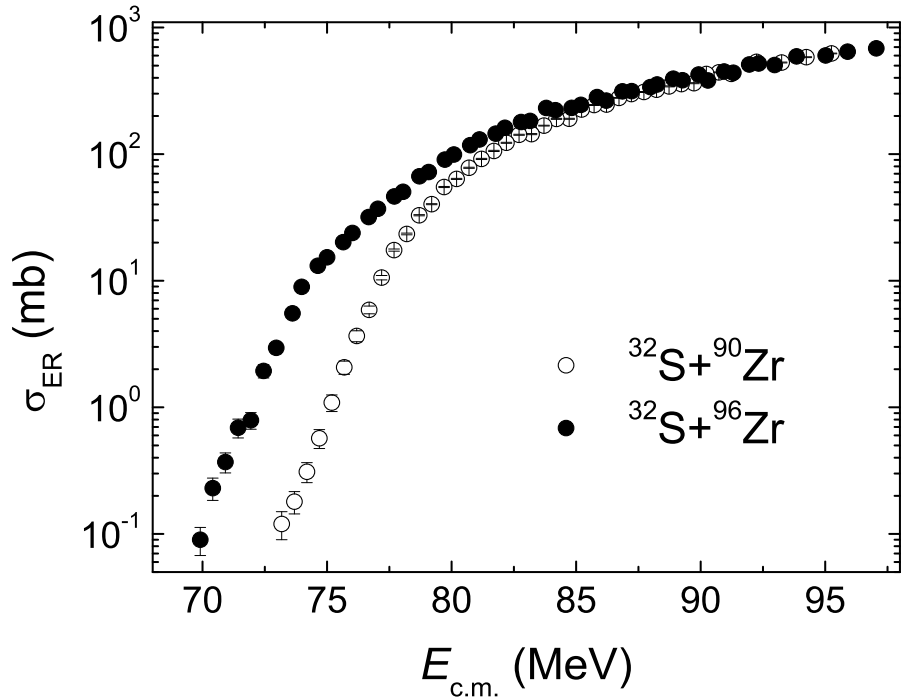} }
\caption{Comparison between the fusion-evaporation (ER) excitation 
  functions of $^{32}$S+$^{90}$Zr (open circles) and $^{32}$S+$^{96}$Zr 
  (points) as a function of the center-of-mass energy. The error
  barrs of the experimental data taken from Ref.~\cite{Zhang10} represent
  purely statistics uncertainties.}
\label{fig:1}       
\end{figure}

In order to investigate the role of neutron transfers we further study 
$^{32}$S + $^{90}$Zr and $^{32}$S + $^{96}$Zr as benchmark 
reactions. Fig.~1 displays the measured fusion cross sections for 
$^{32}$S + $^{90}$Zr (open circles) and $^{32}$S + $^{96}$Zr 
(points). We present the analysis of excitation functions of evaporation 
residues (ER) cross sections recently measured with high precision (i.e. 
with small energy steps and good statistical accuracy for these 
reactions \cite{Zhang10}).

The differential cross sections of quasi-elastic scattering (QEL) at 
backward angles were previously measured by the CIAE group \cite{Yang08}. The
analysis of the corresponding BD-QEL barrier distributions (see solid points 
in Fig.~2) already indicated the significant role played by 
neutron tranfers in the fusion processes.

In Fig.~2 we introduce the experimental fusion-barrier (BD-Fusion) 
distributions (see open poins) obtained for the two reactions by using the 
three-point difference method of Ref. \cite{Rowley92} as applied to the data 
points of Ref. \cite{Zhang10} plotted in Fig.~1. It is interesting to note 
that in both cases the BD-Fusion and BD-QEL barrier distributions are almost 
identical up to E$_{c.m.}$ $\approx$ 85 MeV.

\section{Coupled channel analysis}
\label{sec:2}

We have developed a new CC computer code named NTFus \cite{NTFus} by taking the 
neutron transfer channels into account in the framework of the semiclassical 
model of Zagrebaev \cite{Zagrebaev03}. We will show that the effect of the 
neutron transfer channels yields a fairly good agreement with the present 
data of sub-barrier fusion cross sections measured for $^{32}$S + $^{96}$Zr, 
the more neutron-rich reaction~\cite{Zhang10}. This was initially expected 
from the positive Q-values of the neutron transfers as well as from the failure 
of standard CC calculation of quasi-elastic barrier distributions without 
neutron-transfers coupling \cite{Yang08} as shown by the solid line in 
Fig.~2(b). 

By fitting the present experimental fusion excitation function given in Fig.~1 
with NTFus CC calculation \cite{NTFus}, we will be able to conclude that the 
effect of the neutron transfer channels produces the rather significant 
enhancement of the sub-barrier fusion cross sections of
$^{32}$S + $^{96}$Zr as compared to $^{32}$S + $^{90}$Zr.

\begin{figure}
\resizebox{0.95\columnwidth}{!}{%
\includegraphics{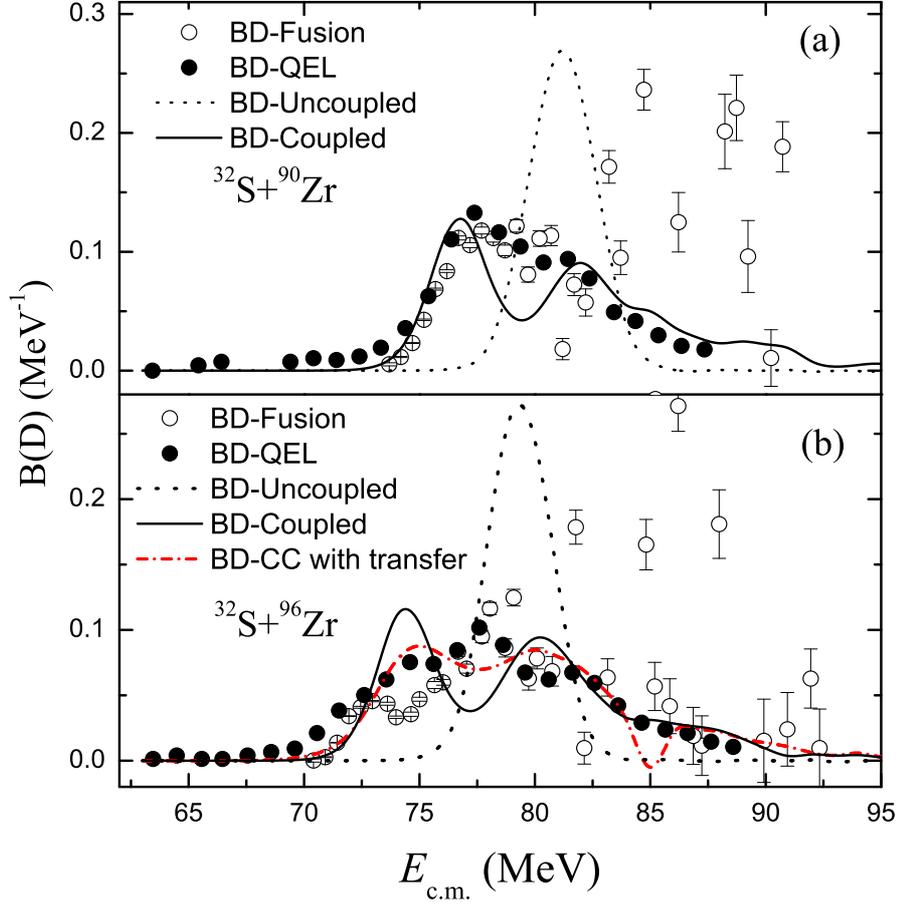} }
\caption{Barrier distributions (BD) from the fusion ER (open circles) 
cross sections \cite{Zhang10}, plotted in Fig.1, and quasielastic scattering 
(solid circles) cross sections \cite{Yang08} for $^{32}$S+$^{90}$Zr (a)
and $^{32}$S+$^{96}$Zr (b). The dashed and solid black lines represent
uncoupled calculations (1D-BPM) and the CC calculations without neutron
transfer coupling. The red dash-dotted line represents the CC calculations
with neutron transfer coupling for the $^{32}$S+$^{96}$Zr reaction.}
\label{fig:2}       
\end{figure}

A detailed inspection of the $^{32}$S + $^{90}$Zr fusion data presented in 
Fig.~1 along with the negative Q-values of their corresponding neutron 
transfer channels lead us to speculate with the absence of a neutron 
transfer effect on the sub-barrier fusion for this reaction. We proceed 
step by step by performing calculations for this reaction with the 
\textsc{NTFus} code \cite{NTFus} (see Fig.~3).

\section{\textsc{NTFus} CC calculations for $^{32}$S+$^{90,96}$Zr}

With the semiclassical model developed by Zagrebaev~\cite{Zagrebaev03} we 
propose in the following discussion to definitively demonstrate the 
significant role of neutron transfers for the $^{32}$S + $^{96}$Zr fusion 
reaction by fitting its experimental excitation function with  
\textsc{NTFus} code \cite{NTFus} calculations, as shown in Fig.~4. The main 
characteristics of the code are briefly described thereafter. 

The new oriented object \textsc{NTFus} code \cite{NTFus}, using the Zagrebaev 
model \cite{Zagrebaev03} was implemented (at the CIAE) in C++, using the 
compiler of \textsc{ROOT} \cite{ROOT}, following the basic equations
of Ref.~\cite{Zagrebaev04}. 

Let us first remind the values chosen for the deformation parameters and 
the excitation energies that are given in Table 1 
\cite{Dasgupta,Raman01,Kebedi05}. The quadrupole vibrations of both
the $^{90}$Zr and $^{96}$Zr are weak in energy; they lie at comparable
energies.

\begin{table}[!h]
\caption{\label{tab1}Excitation energies E$_{x}$, spins and parities 
$\lambda^{\pi}$ and deformation parameters $\beta_{\lambda}$ from 
\cite{Dasgupta,Raman01,Kebedi05}.}
\begin{tabular}{cccc}
Nucleus & E$_{x}$(MeV) & $\lambda^{\pi}$ & $\beta_{\lambda}$\\
\hline
\\
$^{32}$S & 2.230 & 2$^{+}$ & 0.32 \\
 & 5.006 & 3$^{-}$ & 0.40 \\
$^{90}$Zr & 2.186 & 2$^{+}$ & 0.09 \\
 & 2.748 & 3$^{-}$ & 0.22 \\
$^{96}$Zr & 1.751 & 2$^{+}$ & 0.08 \\
 & 1.897 & 3$^{-}$ & 0.27 \\
\end{tabular}
\end{table}

The $^{96}$Zr nucleus presents a complicated situation~\cite{Szilner11}: 
its low-energy spectrum is dominated by a 2$^{+}$ state at 1.748 MeV and 
by a very collective [B(E3;3$^-$ $\rightarrow$ 0$^+$) = 51 W.u.] 3$^-$ 
state at 1.897 MeV. CC calculations explained the larger sub-barrier
enhancement as due mainly to the strong octupole vibration of the 3$^-$
state in $^{36}$S + $^{96}$Zr \cite{Stefanini00}. 
However, the agreement is not so satisfactory below the barrier 
for $^{32}$S + $^{96}$Zr (see solid line of Fig.~4), as well
as  for  $^{40}$Ca + $^{96}$Zr \cite{Timmers98} and, therefore, there is 
the need to take neutron transfers into account.

The main functions of the code \textsc{NTFus} are designed to calculate the 
fusion excitation functions with normalized barrier distribution 
(based on experimental data) given by CCFULL \cite{CCFULL}, we take the 
dynamical deformations into account. 

To take into account the neutron transfers, the \textsc{NTFus} 
code \cite{NTFus} applies the Zagrebaev model~\cite{Zagrebaev03} to 
calculate the fusion cross sections $\sigma_{fus}(E)$ as a function of 
center-of-mass energy E. Then the fusion excitation function can be derived 
using the following formula~\cite{Zagrebaev03}:\\

T$_{l}$(E)~=
 
\begin{equation}
    \int f(B)\frac{1}{N_{tr}}\sum_{k}\int^{Q_{0}(k)}_{-E}\alpha_{k}(E,l,Q) \\
    \times P_{HW}(B,E+Q,l)dQdB
  \label{eq1}.
\end{equation}

and

\begin{equation}
  \sigma_{fus}(E)=\frac{\pi \hbar^{2}}{2\mu E} \sum^{l_{cr}}_{l=0} (2l+1)T_{l}(E)
\label{eq2}.
\end{equation}

where $T_{l}(E)$ are the transmission coefficients, $E$ is the energy given in 
the center-of-mass system, B and $f(B)$ are the barrier height and the 
normalized barrier distribution function, P$_{HW}$ is the usual Hill-Wheeler
formula. $l$ is the angular momentum whereas 
$l_{cr}$ is the critical angular momentum as calculated by assuming no 
coupling (well above the barrier). 
$\alpha_{k}(E,l,Q)$ and $ Q_{0}(k)$ are, respectively, the probabilities and 
the Q-values for the transfers of $k$ neutrons. And $1/N_{tr}$ is the 
normalization of the total probability taking into account the neutron 
transfers. 

The \textsc{NTFus} code \cite{NTFus} uses the ion-ion potential between two 
deformed nuclei as developped by Zagrebaev and Samarin in 
Ref.~\cite{Zagrebaev04}. Either the standard Woods-Saxon form of the nuclear 
potential or a proximity potential \cite{Blocki77} can be chosen. The code
is also able to predict fusion cross sections for reactions induced by halo 
projectiles; for
instance $^{6}$He + $^{64}$Zn \cite{Fisichella11}. In the following, only
comparisons for $^{32}$S + $^{90}$Zr and $^{32}$S + $^{96}$Zr are discussed.

For the high-energy part of the $^{32}$S + $^{90}$Zr excitation function, 
one can notice a 
small over-estimation of the fusion cross sections at energies above the 
barrier up to the point used to calculate the critical angular momentum. 
This behavior can be observed at rather high incident energies - i.e. 
between about 82 MeV and 90 MeV (shown in Fig. 3 for $^{32}$S + $^{90}$Zr 
reaction). We want to stress that the corrections do not affect our 
conclusions that the transfer channels have a predominant role below the 
barrier for $^{32}$S + $^{96}$Zr reaction, as shown in Fig.~4. 

As expected, we obtain a good agreement with calculations not taking any 
neutron transfer coupling into account for $^{32}$S + $^{90}$Zr as shown
by the solid line of Fig.~3 (the dashed line are the results of calculations
performed without any coupling). On the other hand, there is no 
significant over-estimation at sub-barrier energies. As a consequence, it 
is possible to observe the strong effect of neutron transfers on the 
fusion for the $^{32}$S + $^{96}$Zr reaction at sub-barrier energies. 
Moreover, the barrier distribution function $f(B)$ extracted from the data 
contains the information of the neutron transfers. These information are also 
contained in the transmission coefficients, which are the most 
important parameters for the fusion cross sections to be calculated 
accurately. The $f(B)$ function as calculated with the three-point formula 
\cite{Rowley92} will mimic the differences induced by the neutron transfer 
taking place in sub-barrier energies where the cross section variations are 
very small (only visible if a logarithm scale is employed for the fusion 
excitation function). It is interesting to note that the Zagrebaev 
model \cite{Zagrebaev03} implies a modification of the Hill-Wheeler 
probability and does not concern the barrier distribution function $f(B)$.
Finally, the code allows us to perform each calculation by taking the 
neutron transfers into account or not.

\begin{figure}
\resizebox{0.95\columnwidth}{!}{%
\includegraphics{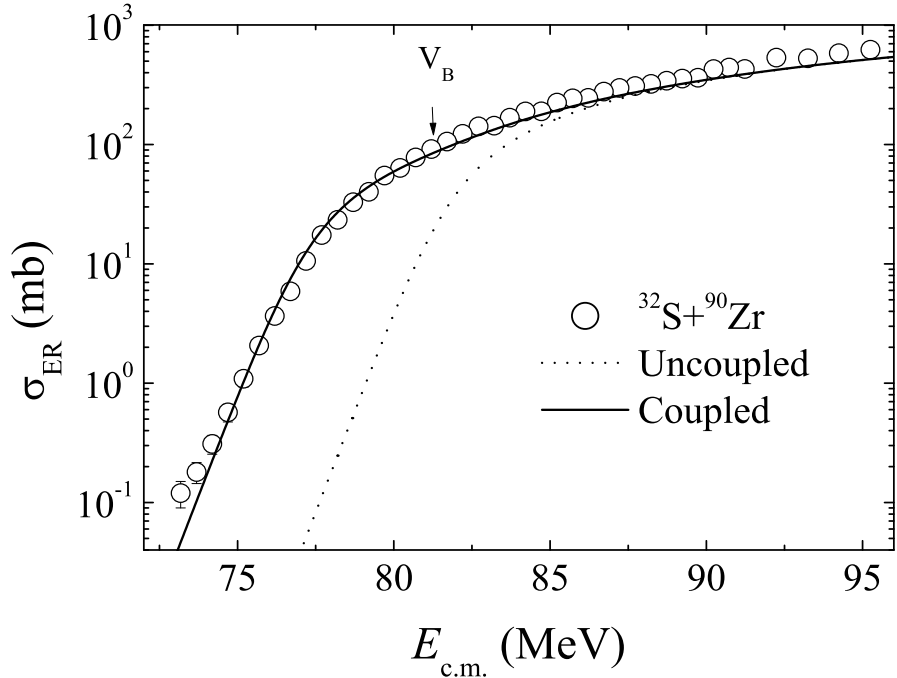} }
\caption{Fusion-evaporation (ER) excitation function for 
$^{32}$S+$^{90}$Zr. The solid points are the experimental data 
\cite{Zhang10} (see Fig.1). The dashed and solid lines 
are the uncoupled calculations, and CC calculations without
neutron transfers, respectively. The arrow indicates the position
of the Coulomb barrier for $^{32}$S+$^{90}$Zr as given by the 
1D-BPM model (see Fig. 2). }
\label{fig:3}       
\end{figure}

\begin{figure}
\resizebox{0.95\columnwidth}{!}{%
\includegraphics{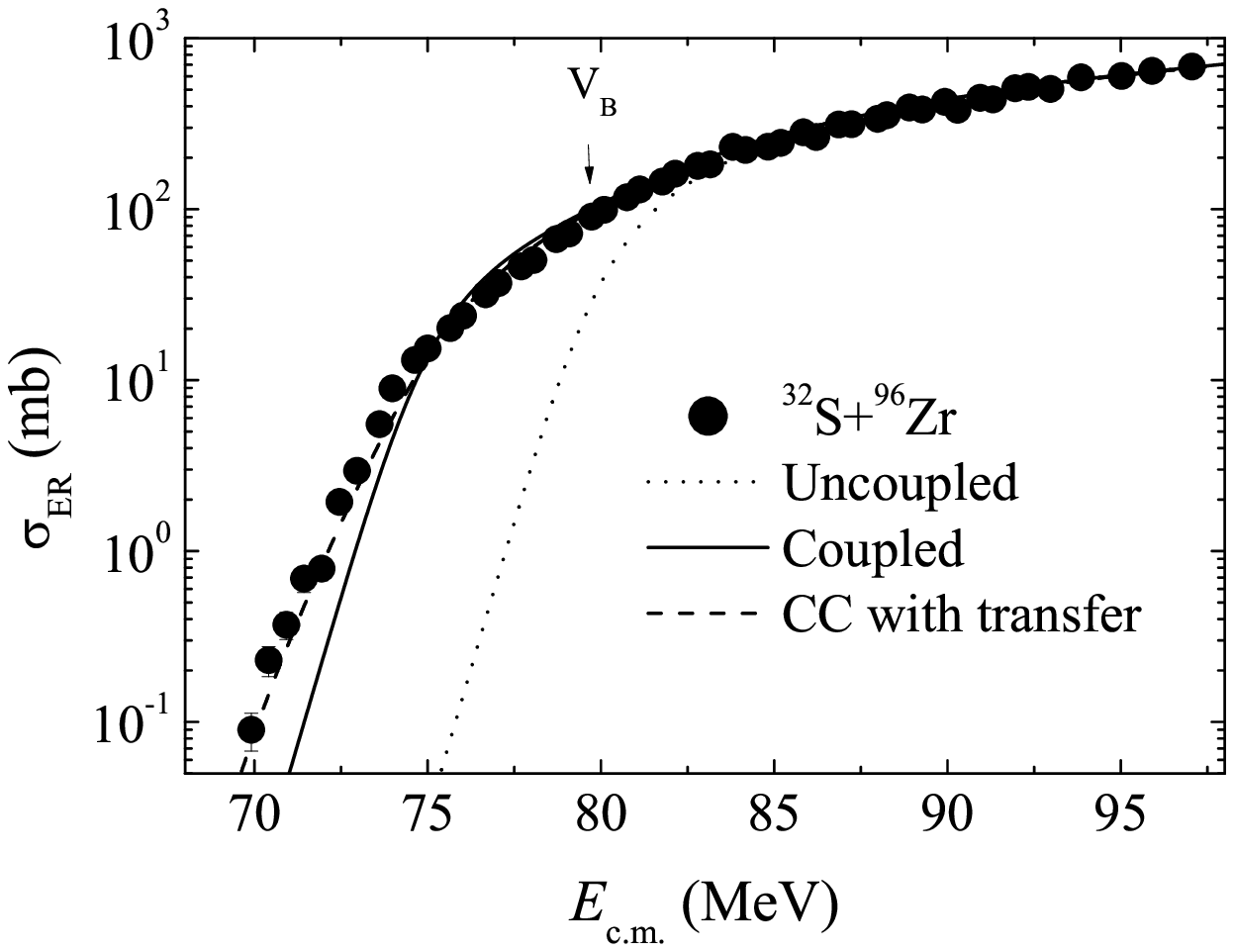} }
\caption{Fusion-evaporation (ER) excitation function for 
$^{32}$S+$^{96}$Zr. The solid points are the experimental data 
\cite{Zhang10} (see Fig.1). The dashed, solid, and dotted lines 
are the uncoupled calculations, and CC calculations without and
with neutron transfers, respectively. The arrow indicates the position
of the Coulomb barrier for $^{32}$S+$^{96}$Zr as given by the 1D-BPM 
model (see Fig. 2).}
\label{fig:4}       
\end{figure}

\begin{table}[!h]
\caption{\label{tab2}Q-values in MeV for neutron pickup transfer channels
from ground state to ground state for $^{32}$S+$^{90}$Zr and $^{32}$S+$^{96}$Zr,
respectevly.}
\begin{tabular}{ccccc}
System & +1n & +2n & +3n & +4n\\
\hline
\\
$^{32}$S+$^{90}$Zr & -3.33 & -1.229 & -6.59 & -6.319 \\
$^{32}$S+$^{96}$Zr & 0.788 & 5.737 & 4.508 & 7.655 \\
\end{tabular}
\end{table}

\begin{table}[!h]
\caption{\label{tab3}Separation energies in MeV of each neutron for 
$^{96}$Zr.}
\begin{tabular}{cccc}
1$^{st}$ neutron & 2$^{nd}$ neutron & 3$^{rd}$ neutron & 4$^{th}$ neutron\\
\hline
\\
 7.854 & 6.463 & 8.230 & 6.733 \\
\end{tabular}
\end{table}

The calculation with the neutron transfer effect is performed here up to the 
channel +4n (k=4), but we have seen that we obtain the same overall agreement
with data up to channels +5n and +6n. The Q-values and the separation energies 
for the $^{96}$Zr nucleus used for this calculation (solid line in Fig. 4) 
are displayed in the Tables 2 and 3, respectively.

As we can see on Fig. 4, the solid line representing standard CC calculations
without the neutron transfer coupling (the dotted line is given for uncoupled
calculations) does not fit the experimental data 
well at sub-barrier energies. On the other hand, the dotted line displaying 
NTFus calculations taking the neutron transfer coupling into account agrees
perfectly well with the data. As expected, the Zagrebaev semiclassical model's 
correction applied at sub-barrier energies enhances the calcutated cross 
sections. Moreover, it allows to fit the data reasonably well and therefore 
illustrates the strong effect of neutron transfers for the fusion of 
$^{32}$S + $^{96}$Zr at subbarrier energies. 
 
The present full CC analysis of $^{32}$S + $^{96}$Zr fusion data 
\cite{Zhang10} using NTFus \cite{NTFus} confirms perfectly well first 
previous CC calculations \cite{Zagrebaev03} describing well the earlier 
$^{40}$Ca + $^{90,96}$Zr fusion data \cite{Timmers98} and, secondly, very 
recent fragment-$\gamma$ coincidences measured for $^{40}$Ca + $^{96}$Zr 
multi-neutron transfer channels \cite{Szilner11}.

\section{Summary and conclusions}

We have investigated the fusion process (excitation functions and extracted 
barrier distributions \cite{Zhang10}) at near- and sub-barrier energies for 
the two neighbouring reactions $^{32}$S + $^{90}$Zr and $^{32}$S + $^{96}$Zr. 
For this purpose a new computer code named NTFus \cite{NTFus} has been 
developped by taking the coupling of the multi-neutron transfer channels 
into account by using the semiclassical model of Zagrebaev \cite{Zagrebaev03}. 

The effect of neutron couplings provides a fair agreement with the present 
data of sub-barrier fusion for $^{32}$S + $^{96}$Zr. This was initially 
expected from the positive Q-values of the neutron transfers as well as from 
the failure of previous CC calculation of quasi-elastic barrier distributions 
without coupling of the neutron transfers \cite{Yang08}. With the agreement 
obtained by fitting the present experimental fusion excitation function
and the CC calculation at sub-barrier energies, we conclude that the effect 
of the neutron transfers produces a rather significant enhancement of the 
sub-barrier fusion cross sections of $^{32}$S + $^{96}$Zr as compared to 
$^{32}$S + $^{90}$Zr. At this point we did not try to reproduce the details 
of the fine structures observed in the fusion barrier distributions. 
We believe that to achieve this final goal it will
first be necessay to measure the neutron transfer cross sections to
provide more information on the coupling strenght of neutron transfer
because its connection with fusion is not yet fully understood \cite{Szilner11}. 

\section{Acknowledgments}

This work was supported by the National Natural Science Foundation of China
under Grants No.~10575134, No. 
10675169, and No.~10735100, and the Major 
State Basic Research Developing Program under a 2007 Grant No.
CB815003. One 
of us (A.R.) thanks Region Alsace of France for the Boussole Grant 2009 
No. 080105519 that was proposed to him during his six-month stay in 2009 at 
the CIAE, in Beijing, China.

\end{document}